\documentclass[fleqn,10pt]{wlscirep}

\usepackage{booktabs}
\usepackage{enumitem}
\usepackage{multirow}
\usepackage{array}
\usepackage{tikz}
\usepackage{subcaption}
\usetikzlibrary{shapes, arrows, intersections, decorations.pathreplacing, decorations.markings}

\usepackage[utf8]{inputenc}
\usepackage[T1]{fontenc}
\usepackage{lineno}

\title{Why we should respect analysis results as data}

\author[1,*]{Joana M Barros}
\author[1]{Lukas A Widmer}
\author[1*]{Mark Baillie}
\author[1]{Simon Wandel}
\affil[1]{Analytics, Novartis Pharma AG, Basel, Switzerland}

\begin{abstract}
The development and approval of new treatments generates large volumes of results, such as summaries of efficacy and safety. However, it is commonly overlooked that analyzing clinical study data also produces data in the form of results. For example, descriptive statistics and model predictions are data. Although integrating and putting findings into context is a cornerstone of scientific work, analysis results are often neglected as a data source. Results end up stored as ``data products'' such as PDF documents that are not machine readable or amenable to future analysis. We propose a solution to ``calculate once, use many times'' by combining analysis results standards with a common data model. This analysis results data model re-frames the target of analyses from static representations of the results (e.g., tables and figures) to a data model with applications in various contexts, including knowledge discovery. Further, we provide a working proof of concept detailing how to approach analyses standardization and construct a schema to store and query analysis results. 
\end{abstract}

\begin{document}

\flushbottom
\maketitle

\thispagestyle{empty}


\section*{Introduction}
The process of analyzing data also produces data in the form of results. In other words, project outcomes themselves are a data source for future research: aggregated summaries, descriptive statistics, model estimates, predictions, and evaluation measurements may be reused for secondary purposes. For example, the development and approval of new treatments generates large volumes of results, such as summaries of efficacy and safety from supporting clinical trials through the development phases. Integrating these findings forms the evidence base for efficacy and safety review for new treatments under consideration.   
\par
Although integrating and putting scientific findings into context is a cornerstone of scientific work, project results are often neglected or indeed not handled as data. Analysis results are typically shared as part of presentations, reports, or publications addressing a greater objective. The results of data analysis end up stored as \textit{data products}, namely, presentation-suitable formats such as PDF, PowerPoint, or HTML documents populated with text, tables, and figures showcasing the results of a single analysis or an assembly of analyses. Data products are not designed to be machine-readable or amenable to future data analysis. 
This is the case for clinical trial reporting where the data analysis summaries from a study are rendered to rich text format (RTF) files that are then compiled into appendices following the International Council for Harmonisation of Technical Requirements for Pharmaceuticals for Human Use (ICH) E3 guideline \cite{emea1996} where each appendix is a table, listing or and figure summary of a drug efficacy and safety evaluation. The analysis results stored in these appendices - which can span 1000s of pages - are not readily reusable without digitisation. 
There have been recent attempts to modernise the reporting of clinical trials including the use of electronic notebooks and web-based frameworks. However, while literate programming documents such as Rmarkdown allow documenting code and results together and R-shiny enables dynamic data exploration, the rendered data products also suffer the same fate of presentation-suitable formats. In other words, modern data products also do not handle data analysis results as data. 
\par
A focus on results presentation over storage considerations sets up a barrier impeding the assimilation of scientific knowledge, understanding what was intended and what was implemented. As a repercussion, the scientific process cycle is broken, leaving researchers who want to reuse prior results with three options:
\begin{enumerate} 
  \item Re-run the analysis if the code and original source data are accessible.
  \item Re-do the analysis if only the original source data is accessible.
  \item Manually or (pseudo-)automatically extract information from the data products (e.g., tables, figures, published notebooks).
\end{enumerate}
The first option would appear to be the best one and is, for instance, being implemented in Elife executable research articles\cite{elife}. However, being able to rerun the analysis does not guarantee reproducibility and can be computationally expensive when covering many studies, large data, or sophisticated models. Analyses can depend on technical factors such as the products used, their versions, and (hardware and software) dependencies, all of which affect the outcome. Even tailored statistical environments such as R \cite{r} have a wide range of output discrepancies and must rely on extensions, such as \textit{broom} \cite{broom} for reformatting and standardizing the outputs of data analysis. 
\par
For the second option, there are additional complications to account for: even if we assume that the entire analysis is fully documented, common analyses are not straightforward to implement. This option assumes that the complete details required to implement the analysis are documented, for example, in a statistical analysis plan (SAP). In our experience however, data-driven and expertise-driven undocumented choices are a hidden source of deviations that make reproducing or replicating the results an elusive task. On top of this, the selective reporting of results limits replication of the complete set of performed data analyses (both pre-specified and ad-hoc) within a research project \cite{gelman2013garden,wicherts2016degrees,devezer2020case}. 
\par
The last scenario is common place for secondary research that combines and integrates findings of single, independent studies, such as meta-analyses or systematic reviews. To assess the findings it is necessary to first digitize the studies' documents either through a laborious manual effort or by using extraction tools known to be error-prone and requiring verification. Furthermore, the unavailability of complete results, potentially through selective reporting, requires researchers to extrapolate the missing results, which can lead to questionable reliability and risk of bias \cite{tendal2009disagreements}.
\par
Data management is an important, but often undervalued, pillar of scientific work. Good data management supports key activities from planning and execution to analysis and reporting. The importance of data stewardship is now also recognized as an additional pillar. Good data stewardship supports activities beyond the single project into areas such as knowledge discovery, as well as the reuse of data for secondary purposes, to other downstream tasks such as the contextualization, appraisal, and integration of knowledge. Initiatives like FAIR set up the minimal guiding principles and practices for data stewardship based on making the data Findable, Accessible, Interoperable, and Reusable \cite{wilkinson2016fair}. Likewise, the software and data mining community have introduced initiatives bringing standardization to analytic applications, thus facilitating data exchange and releasing the researcher from the burden of translating the output of statistical analysis into a suitable format for the data product \cite{grossman1999management,pfa2022,onnx,ibmmodel}. 

\begin{figure}
\centering
\includegraphics[width = \linewidth]{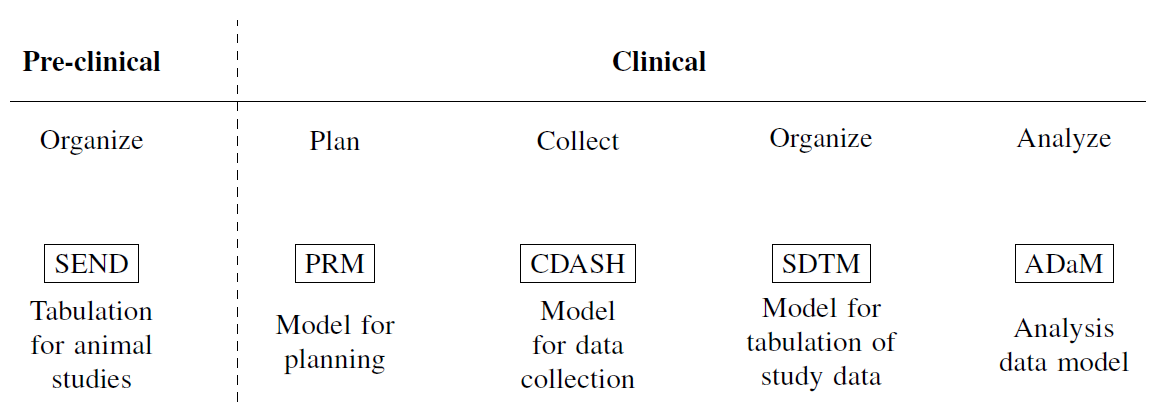}
\caption{CDISC defines a collection of standards adapted to the different stages in the clinical research process. For example, ADaM defines data sets that support efficient generation, replication, and review of analyses \cite{cdisc2020standardsresearch}.}
\label{fig:cdisc_standards}
\end{figure}

An important component of data management is the data model which specifies the information to capture, how to store it, and standardizes how the elements relate to one another. In the clinical domain, data management is a critical element in preparing regulatory submissions and to obtain market approval. In 1999 the Clinical Data Interchange Standards Consortium (CDISC) introduced the operational data model (ODM) facilitating the collection, organization, and sharing of clinical research data and metadata \cite{huser2015standardizing}. In addition, the ODM enabled the creation of standards (figure~\ref{fig:cdisc_standards}) such as the Standard Data Tabulation Model (SDTM) and the analysis data model (ADaM) to easily derive analysis datasets for regulatory submissions \cite{cdiscsadam}. However, CDISC data standards only consider data from planning and collection, up to analysis data (i.e. data prepared and ready for data analysis). 
\par
In this paper, we explore the concept of viewing the output of data analysis as data. By doing so, we address the problems associated with the limited reproducibility and reusability of analysis results. We demonstrate why we should respect analysis results as data and put forward a solution using an \textit{analysis result data model} (ARDM), re-framing the analyses target from the applications of the results (e.g., tables and figures) to a data model. By integrating the analysis results into a similar schema with specific constraints, we would ensure data quality, improve reusability, and facilitate the development of tools leveraging the re-use of analysis results. Taking meta-analyses again as an example, applying an ARDM would now only require one database query instead of a long process of information extraction and verification. Tables, listings, and figures could be generated directly from the results instead of repeating the analysis. Furthermore, storing the results as independent datasets would also allow sharing information without the need for the underlying individual patient data, a useful property given data protection regulations in both academic and industry publications. Viewing analysis results as a data source moves us from repeating or redundantly recording results to a \textit{calculate once, use many times} mindset.


\section*{The analysis results data model}
To create an analysis results data model, the first step requires thinking of the results of the analysis as data itself. Through this abstraction, we can begin organizing the data in a common model linking (e.g., clinical) datasets with the analysis results. Before we further introduce the ARDM it is necessary to clarify what an analysis and analysis results entail. An analysis is formally defined as a ``detailed examination of the elements or structure of something’’ \cite{analysiscambridge}. In practice, it is a collection of steps to inspect and understand data, explore a hypothesis, generate results, inferences, and possibly predictions. Analyses are fluid and can change depending on the conclusions drawn after each one of the steps. Nonetheless, routine analyses promote conventions that we can use as a foundation to create analysis standards. For example, looking at the table of contents of a Clinical Study Report (CSR) we can see a collection of routine results summaries. Diving deeper into these sections, we can see the same or similar analysis results between CSRs of independent clinical studies, namely due to conventions \cite{emea1996}. For example, it is standard for a clinical trial to report the demographics and baseline characteristics of the study population, and a summary of adverse events. These data summaries, may also be a collection of separate data analyses grouped together in tables or figures (i.e., descriptive statistics of various baseline measurements, or the incidence rates of common adverse drug reactions, by assigned treatment). Also, the same \textit{statistics}, such as the number of patients assigned to a treatment arm, may be repeated throughout the CSR. Complex inferential statistics may also be repeated in various tables and figures. For example, key outcomes maybe grouped together in a standalone summary of a drug's benefit-risk profile. Therefore, without upfront planning, the same \textit{statistics} may be implemented many times in separate code. 

\begin{figure}
\centering
\includegraphics[width = \linewidth]{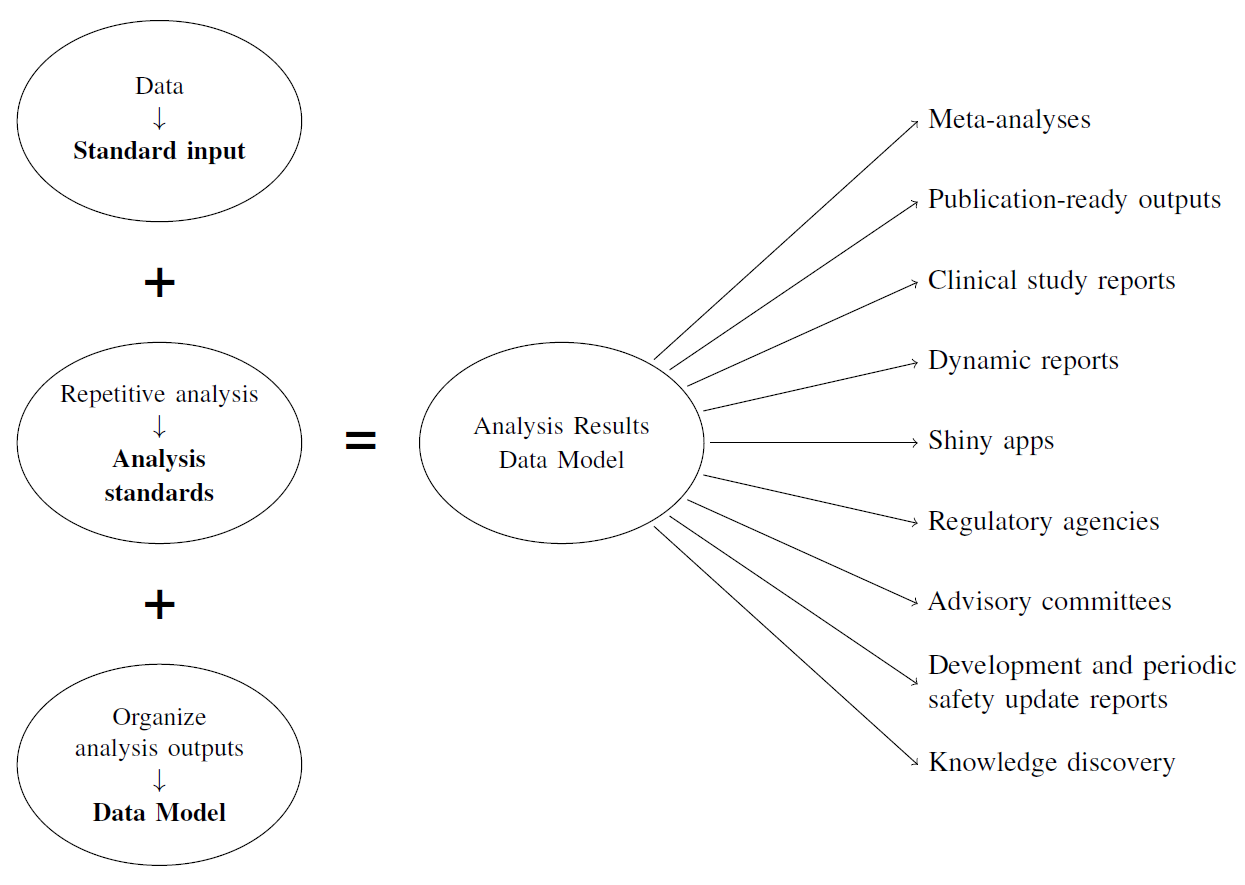}
\caption{In clinical development, the analysis results data model enables a source of truth for results applied in various applications. Currently, the examples on the right require running analyses independently, even when using the same results.}
\label{fig:applications}
\end{figure}

The analysis results are the outcome of the analysis and are typically rendered into tables, figures, and listings to facilitate the presentation to stakeholders. Some examples of applications that can reuse the same results are present in figure~\ref{fig:applications} (right). Before the rendering, the results are stored in intermediate formats such as data frames or datasets. We can use this to our advantage and capture the results for posterior use in research by defining which elements to store and the respective constraints. This supports planning the analyses and the potential applications for the results, minimizing imprudent applications. An analysis results data model can be used to formalize the result elements to store and the constraints with the additional benefit of making the relationships between the results explicit. For example, we can store intermediate results, generated after the initial analysis steps, and use them to achieve the final analysis results. Besides improving the reusability of results, and reproducibility of the analysis, establishing relationships enables retracing the analysis steps and promotes transparency.
\par
Data standards are useful to integrate and represent data correctly by specifying formats, units, and fields, among others. Due to the many requirements in clinical development, guidelines detailing how to implement a data standard are also frequent and essential to ensure the standard is correctly implemented and to describe the fundamental principles that apply to all data \cite{ADaMig}. An \textit{analysis standard} would thus define the inputs and outputs of the analysis as well as the steps necessary to achieve those outputs. While an analysis convention follows a general set of context-dependent analysis steps, a standard ensures the analysis steps are inclusive (i.e., independent of context), consistent and uniform where each step is specified through a grammar \cite{ wilkinson2012grammar,wickham2014tidy,lee2019plyranges} or the querying syntax used in database systems \cite{iso}. 

\begin{figure}
\centering
\includegraphics[width = \linewidth]{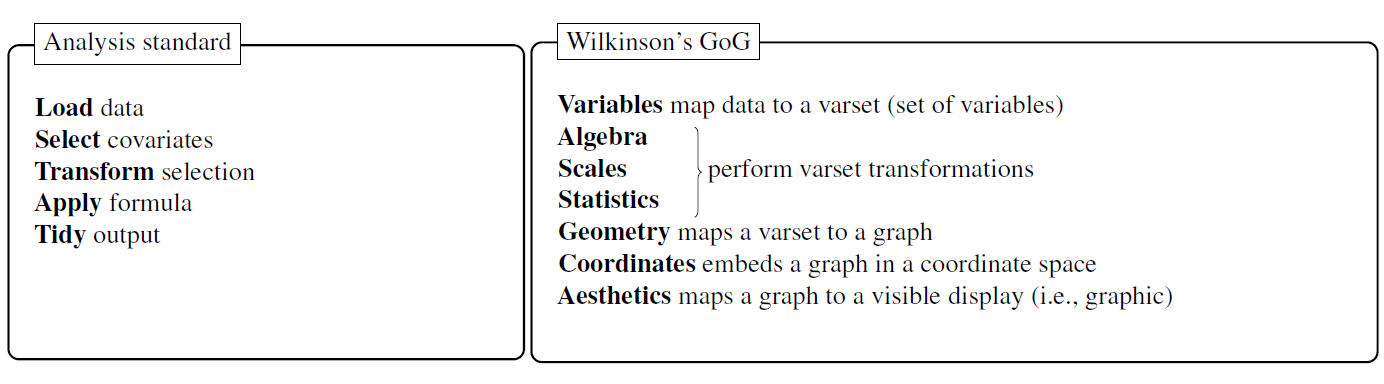}
\caption{The analysis standard follows a grammar to define the steps in the analysis. Similarly,  Wilkinson's\cite{wilkinson2012grammar} grammar of graphics (GoG) concisely defines the components to produce a graphic.}
\label{fig:analysis}
\end{figure}

In figure~\ref{fig:analysis}, we compare the concepts behind an analysis standard with Wilkinson’s grammar of graphics (GoG) data flow. Both follow an immutable order, ensuring that previous steps must be fulfilled to achieve the end result. For example, any data transformation needs to occur before we apply a formula (e.g., compute the descriptive statistics), otherwise, the result of the analysis becomes dubious. The collection of steps forms a grammar; however, each step also offers choices. For example, \textit{apply formula} can refer to a linear model or Cox model. Wilkinson refers to this characteristic as the system's richness by the means of ``paths” constructed by choosing different designs, scales, statistical methods, geometries, coordinate systems, and aesthetics. In the context of the ARDM, analysis standards support pre-planning, compelling the researcher to iterate over the potential analysis routes and the underlying question the analysis should address. In general, it is good practice to write down the details of an analysis, for example using a SAP, with sufficient granularity that the analysis could be reproduced independently if only the source data was available. Thus, the analysis standards would translate the intent expressed in the SAP into clear and well-defined steps. The immediate benefits are transparency and reproducibility with future gains in automation. In clinical development, standard operating procedures already cover many of these steps. However, they critically do not handle analysis results as a data source. Combining a data model with analysis standards would benefit clinical practice in four aspects:
\begin{enumerate}
\item Guaranteeing data quality and consistency across a clinical program, essentially creating a single source of truth designed to handle different levels of project abstraction. For example, from a single data analysis to a complete study, or a collection of studies. 
\item Reusability by providing standardization across therapeutic areas and instigating the development of tools using the results instead of requiring individual patient data (e.g., interactive apps).
\item Simplicity as the analysis standard would encourage upfront planning and identify the necessary inputs, steps, and outputs to keep (e.g., reducing the complexity of forest plots and benefit-risk graphs summaries).
\item Efficiency by avoiding the manual and recurrent repetition of the analysis, and leveraging modularization and standardization of inferential statistics.
\end{enumerate}
Analysis results datasets have been previously put forward as a solution to improve the uptake of graphics within Novartis, under the banner of graph-ready datasets \cite{Vandemeulebroecke2019}. Experienced study team leads often have implemented it for efficiency gains, especially around analysis outputs that would reuse existing summary statistics, for example, to support benefit-risk graphs where outcomes may come from different domains. Our experience has also revealed an element of institutional inertia. Standardizing analysis and results requires upfront planning which is often seen as added effort. However, teams that have gone through the steps of setting up a data model and a lightweight analysis process, have found efficiency and quality gains in reusing and maintaining code, as well as verifying and validating results. Regarding inferential results, instead of using results documents or repeating an analysis, we can simply access a common database where these are stored. An ARDM also simplifies modifications to the analysis (and consequently the results). With current practice, these changes might impact one function, program, or script in the best case, or multiple programs or scripts in the worst case. Using an ARDM only requires changes to one program as these can automatically propagate to any downstream analyses. The validation is also simplified as we transition from comparing data products (e.g., RTF files and plots) to comparing datasets directly. Additionally, this brings clarity and transparency, and is suitable for automation.

\subsection*{Six guiding principles}
To create the ARDM we follow a collection of principles addressing the obstacles commonly faced during the clinical research process but also present in other areas. These principles are highlighted in table \ref{tab:principles} and broadly put forward improvements to quality, accessibility, efficiency, and reproducibility. On top of providing a data management solution, the ARDM compels us to take a holistic view of the clinical research process, from the initial data capture to the potential end applications. With this view, we have a clearer picture of where deficiencies occur and of their impact on the process.
\par
In many industries where sub-optimal but quick solutions are preferred, technical debt is a growing problem \cite{techdebt}. While some amount of technical debt is inevitable, understanding our processes can point us to where to make progressive updates and improvements. For example, upfront planning using analysis standards would reduce this debt by default as our starting point are previously verified and validated analyses (i.e., analysis standards). In an effort to continue reducing the debt, the ARDM separation of concerns principle streamlines changes and updates to processes since the analysis, results, and products are separate entities. Standardizing how to store results enables the use of different programming languages to perform analysis with traditionally non-comparable output formats (e.g., SAS and R). Furthermore, we believe the ARDM should grow organically and community-driven, supporting consensus building and cross-organization access.

\begin{table}[!ht]
\caption{The analysis results data model (ARDM) follows six principles broadly addressing the challenges in ensuring reproducible, traceable, reusable, and interoperable results.}
\begin{tabular}{m{5cm}p{9.5cm}}
    \toprule
    \textbf{Principle} & \textbf{Justification} \\ \midrule
    Searchable & 
        \begin{itemize}[leftmargin=*, topsep=0pt, noitemsep]
            \item Consistent location to track and find results (i.e., knowledge source).
            \item Support search across results.
            \item Coherent querying through tools such as APIs.
            \item Accessible to data consumers and users.
        \end{itemize}\\ \midrule
    Nonredundant & 
        \begin{itemize}[leftmargin=*, topsep=0pt, noitemsep]
            \item Designed to acknowledge the \textit{grammar} of data analysis  (i.e., \textit{estimate}, \textit{store}, \textit{retrieve}, \textit{render}).
            \item Avoid repetitions (i.e., scripts and results).
            \item Control technical debt.
            \item Establish a single source of truth.
        \end{itemize}\\ \midrule
    Separation of concerns & 
        \begin{itemize}[leftmargin=*, topsep=0pt, noitemsep]
            \item Modularizes workflows to facilitate updates and expansion.
            \item Streamline validation and verification.
        \end{itemize}\\\midrule
    Reusable and extensible & 
        \begin{itemize}[leftmargin=*, topsep=0pt, noitemsep]
            \item Primary use: support ready-to-use outputs (e.g., plots for clinical reports).
            \item Secondary use: reusable for subsequent analyses (e.g., meta-analysis and systematic reviews).
            \item Supports future extensions (e.g., novel outputs and updates to known  applications).
        \end{itemize}\\ \midrule
    Interoperable &  
        \begin{itemize}[leftmargin=*, topsep=0pt, noitemsep]
            \item Results and metadata are stored in a consistent data format.
            \item Results are represented by common or shared vocabularies (i.e., following a data model).
            \item Enables standardization of outputs across programming languages bypassing language-specific conventions.
        \end{itemize}\\ \midrule
    Community-driven &  
        \begin{itemize}[leftmargin=*, topsep=0pt, noitemsep]
            \item Consensus building and accessible within and across organizations.
            \item Standards and tools are open, designed by and for the scientific community.
        \end{itemize}\\
    \bottomrule
\end{tabular}
\label{tab:principles}
\end{table}


\section*{Implementing the ARDM in clinical research}
The ARDM is adaptive and expandable. For example, with each analysis standard, we can adapt or create new tables to the schema. With respect to the inspection and visualization of the results, there is also the flexibility to create a variety of outputs, independent of the analysis standard. The proof of concept for the ARDM is implemented using the R programming language, and a relational SQLite database \cite{sqlite2022hipp}; however, these choices can be revisited as the ARDM can be implemented using a variety of languages and databases. This implementation should be viewed as a starting point rather than a complete solution. Here, we highlight the considerations we took to construct the ARDM utilizing three analysis standards (descriptive statistics, safety, and survival analysis) and leveraging the CDISC Pilot Project ADaM dataset. Further documentation is available in the code repository. An overview of the requirements to create the ARDM is shown is figure~\ref{fig:applications}
\par
Prior to ingesting clinical data, the algorithm first creates empty tables with specifications on the column names and data types. These tables are grouped into metadata, intermediate data, and results. The metadata tables are created to record additional information such as variables types (e.g., categorical and continuous) and measurement units (e.g., age is given in years). As part of the metadata tables, the algorithm also creates an analysis standards table requiring information on the analysis standard name, function calls, and its parameters. The intermediate data tables aggregate information at the subject level and are useful to avoid repeated data transformations (e.g., repeated aggregations) thus, reducing potential errors and computational execution time during the analysis. The results tables specify the analysis results information that will be stored. Note that the creation of the metadata, intermediate data, and result tables require upfront planning to identify which information should be recorded. Although it is possible to create tables ad hoc, a fundamental part of the ARDM is to generalize and remove redundancies rather than creating a multitude of fit-for-purpose solutions. Hence, creating a successful ARDM requires understanding the clinical development pipeline to effectively plan the analysis by taking into account the downstream applications of the results (e.g., the analysis standard or the data products). As the information stored in the results tables is dictated by the data model, it is possible to inspect the results by querying the database and creating visualizations. In the public repository, we showcase how to query the database and create different products from the results. Furthermore, the modular nature of the ARDM separates the results rendering from the downstream outputs hence, updates to the data products do not affect the results.

\subsection*{Applications}
Analysis standards are a fundamental part of the ARDM to guarantee coherent and suitable outputs. They ensure that the results are comparable, which is not always the case. Similarly, where conventions exist (e.g., safety analysis), we can use an ARDM to provide structure to the results storage thus, facilitating access and reusability. In short, it provides a knowledge source of validated analysis results, i.e. a single source of truth. This enables the separation between the analysis and the data products, streamlining the creation of tables or figures for publications, or other products as outlined in figure~\ref{fig:applications}.
\par
Tracking, searching, and retrieving outputs is facilitated by having an ARDM as it enables query-based searches. For example, we can search based on primary endpoints ``p-value'', ``point estimates'', and adverse events incidence for any given trial present in the database. With automation, we can also select cohorts through query-based searches and apply the analysis standards to automate the creation of results using the selected data. This also facilitates decision-making and enhancements. For example, one can have access to complete trial results beyond the primary endpoint, and extrapolate to cohorts that require special considerations such as pediatric patients. In addition, a single source of truth for results encourages the adoption of more sophisticated approaches to gather new inferences, for example, using knowledge graphs and network analysis.

\subsection*{Case study: Updating a Kaplan–Meier plot}
\begin{figure}[!ht]
\centering
\begin{subfigure}{.46\linewidth}
  \centering
  \includegraphics[width = \linewidth]{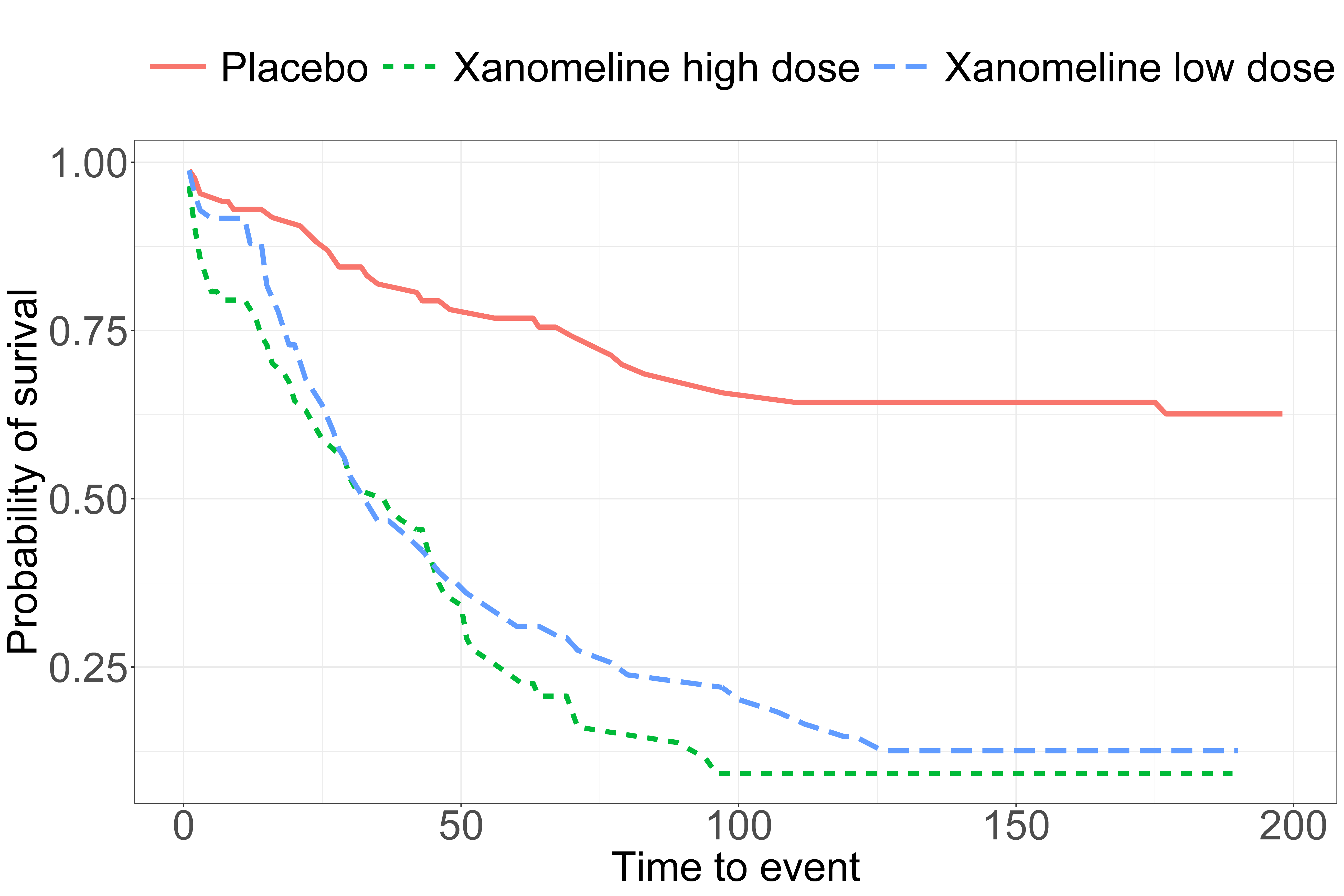}
\caption{Plot with three survival curves.}
  \label{fig:km3}
\end{subfigure}
\hspace{2em}
\begin{subfigure}{.46\linewidth}
  \centering
  \includegraphics[width = \linewidth]{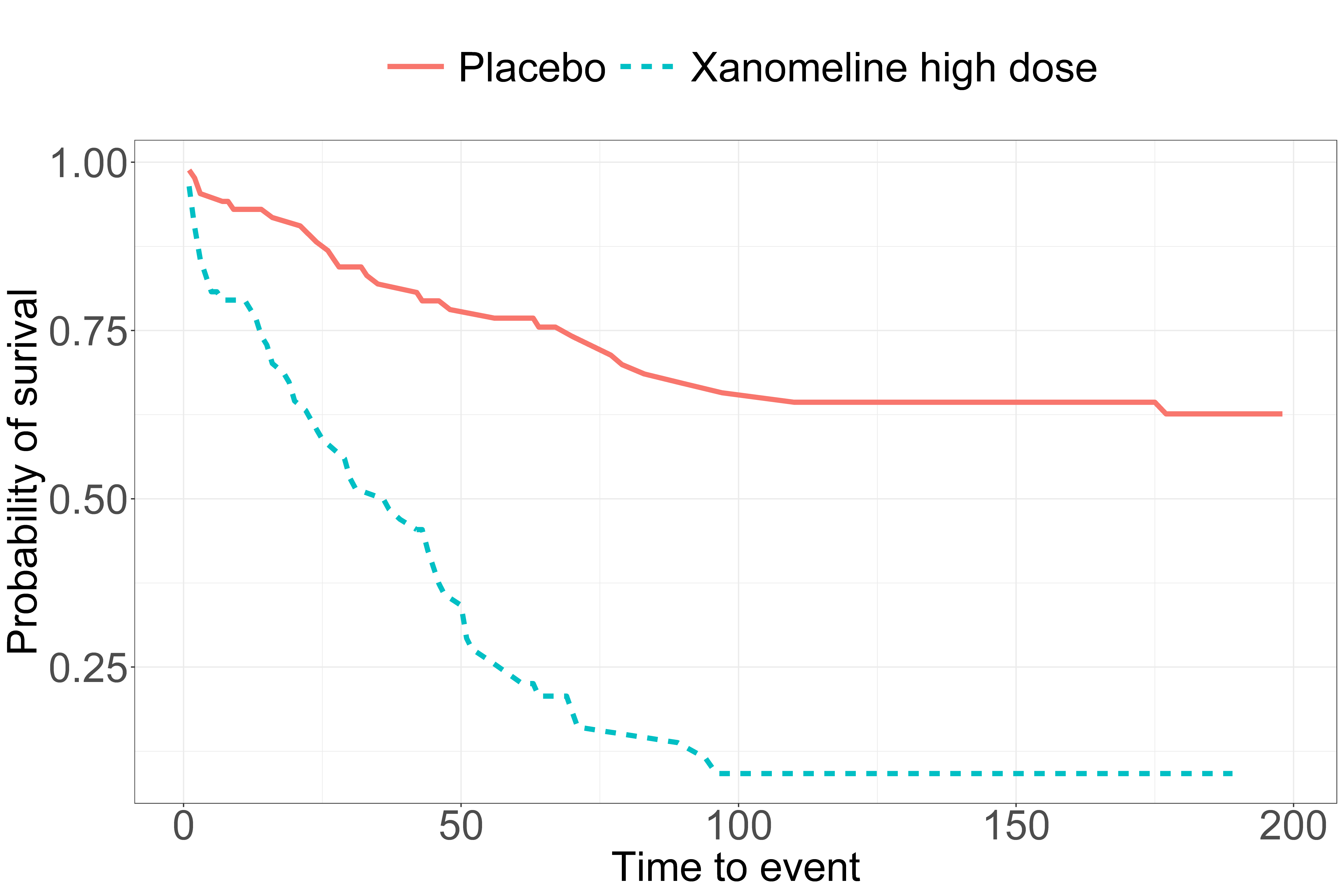}
  \caption{Updated plot with two survival curves.}
  \label{fig:km2}
\end{subfigure}
\caption{A Kaplan-Meier plot example using the CDISC Pilot Project ADaM dataset \cite{cdiscpilotdata}. Without an ARDM, a simple update from \ref{fig:km3} to \ref{fig:km2} requires access to the analysis code and source data which can be a time and resource consuming process. With results stored in an ARDM, the update becomes a simple render to remove a stratum.}
\label{fig:application_km}
\end{figure}

The Kaplan-Meier plot is a common way to visualize the results from a survival or time-to-event analysis. The purpose of the Kaplan-Meier non-parametric method is to estimate the survival probability from observed survival times \cite{kaplanmeier1958}. Note that some patients might not experience the event (e.g., death, relapse); hence, censoring is used to differentiate between the cases and to allow for valid inferences. As a result of the analysis, survival curves are created for the given strata. For example, figure~\ref{fig:km3} shows a Kaplan-Meier plot with three strata corresponding to the treatments in the CDISC pilot study.

\par
A results visualization can appear in a variety of documents from presentation slides, an initial report, or a final publication, however, it is most likely not accompanied by the results used to create it. This hinders the reuse of the information (i.e., results) in the plot. In figure~\ref{fig:km2}, one stratum is removed and the plot only shows two survival curves. This update may seem trivial but, from our experience, this task can require considerable time and effort due to the unavailability of the results. Without an analysis results data model or a known location where to find the results from the survival analysis, one must first locate the clinical data to perform the same analysis again. Then, search for and find the analysis code and the instructions to create the Kaplan-Meier plot. Eventually, one must repeat the analysis entirely. Thirdly, it is advisable to confirm whether the new plot matches the one we want to update; this is especially important if the analysis had to be redone as data transformations might have happened (e.g., different censoring than originally planned). Finally, one can filter the strata and create the plot in figure~\ref{fig:km2}. 
\par
Even in the showcased scenario, we assume to have access to the clinical data, however, this might not be the case. Data protection is an important aspect of any research area. While data protection regulations have provided a way to share data and in return improve the reproducibility of experiments, in clinical research, sharing sensitive subject-specific data is impractical or simply not possible for legal reasons. Another option is to only share aggregated data or the analysis results. While this option can still bring privacy issues, for example due to the presence of outliers, results are already widely shared in publications through visualizations like the ones shown in figure~\ref{fig:application_km}. In contrast to current practice, having an ARDM in place gives many options on what data to share to support results reusability in a variety of contexts. For example, even regulatory agencies can benefit from the ARDM since outputs such as tables, graphics and listings can be easily generated from the results without the need to repeat or reproduce analyses. From our experience, it is common to initially share results with limited people (e.g., within a team) where we do not give much importance to details like aesthetics. However, at a later stage, researchers need the results to update the visualization to suit a wider audience, or use this data for future research. In the Kaplan-Meier plot example, this requires reverse-engineering by using tools to digitize the plot and create machine-readable results.


\section*{Discussion}
The ARDM provides a solution to handle analysis results as data by creating a single source of truth. To guarantee the accuracy of the source, it leverages analysis standards (i.e., validated analyses) with known outputs which are then organized in a database following the proposed data model. The use of analysis standards supports the pre-planning of analyses, compelling the researcher to iterate on the best approach for analyzing the data, and potentially deciding to use pre-existing and appropriate analysis standards.
\par
The concept of creating standards through a common data model is recognised as good data management and stewardship practice. A few examples include the Observational Medical Outcomes Partnership data model, a standard designed to standardize the structure and content of observational data \cite{omopcmd} and the Large-scale Evidence Generation and Evaluation across a Network of Databases research initiative to generate and store evidence from observational data \cite{schuemie2020principles}. The data model created by the Sentinel initiative, led by the Food and Drug Administration (FDA), is tailored to organize medical billing information and electronic health records from a network of health care organizations \cite{sentinelcmd}. Similarly, the National Patient-Centered Clinical Research Network also established a standard to organize the data collected from their network of partners \cite{pcornet}. Finally, expanding the search to translational medicine, the Informatics for Integrating Biology and the Bedside introduced a standard to organize electronic medical records and clinical research data \cite{i2b2datamodel,murphy2010serving}.
\par
Alongside data models, standard processes have been established to generate analysis results such as the requirement to document analyses in SAPs \cite{gamble2017guidelines}, including all data transformations from the source data to analysis ready data sets. However, analyses can be complex and dependent on technical factors, such as the statistical software used, as well as undocumented analysis choices throughout the pipeline, from source data to result. Even less complex routine analyses are error-prone and might not be clearly reproducible. Altogether, this process is time and resource-consuming. A proposed solution is to perform the analysis automatically. With this in mind and targeting clinical development, Brix et al\cite{brix2018odm} introduce the ODM Data Analysis, a tool to automatically validate, monitor, and generate descriptive statistics from clinical data stored in the CDISC Operational Data Model format. The FDA's Sentinel Initiative is also capable of generating descriptive summaries and performing specific analysis leveraging the proprietary Sentinel Routine Querying System \cite{sentinelqueryingsystem}.
\par
Following this direction, the natural progression would be to create a standard suited for storing analysis results. Such an idea is implemented in the genome-wide association studies (GWAS) catalogue where curators assess GWAS literature, extract data, and store it following a standard including the summary statistics \cite{gwassummary}. Taking a step in this direction, CDISC began the 360 initiative to support the implementation of standards as linked metadata in an attempt to improve efficiency, consistency, and reusability across the clinical research \cite{cdisc360}. Nonetheless, the irreproducibility of research results remains an obstacle in clinical research and has brought up calls for global data standardization to enable semantic interoperability and adherence to the FAIR principles \cite{jauregui2019turning}. In our view, the analyses standards and the ARDM are an important contribution to this initiative.
\par
Utilizing the proposed ARDM has a set of requirements. Firstly, the provided clinical data must follow a consistent standard (i.e., CDISC ADaM). Our solution involves automatically populating a database, hence there are expectations regarding the structure of the data. Similarly, data standards are necessary to enable analysis standards. If the analysis input expectations are not met, the analysis is unsuccessful and no results are produced or stored. Further, when a data standard is updated it is necessary to also update the analysis standards and the ARDM accordingly. Another limitation is the necessity of analysis standards. Without quality analysis standards, the quality of the source of truth is not guaranteed. Creating analysis standards requires a good understanding of the analysis to correctly define the underlying grammar and identify relevant decision options for the user (e.g., filter data before modeling). The third limitation corresponds to the applications. At the moment, the ARDM stores and organizes results in a suitable way to reuse in known applications (e.g., creating plots, tables, and requesting individual result values). As future applications are unknown, the data model might not store all the information needed. However, given the ARDM modular approach, it is only necessary to update the result information to be kept rather than updating the entire workflow. Another limitation refers to the supported data modalities. The proposed ARDM is implemented on tabular clinical trial data. However, it is possible to adapt the ARDM and design choices (e.g., type of database) to support diverse data. For example, the summary statistics present in the genome-wide association studies (GWAS) catalog \cite{gwassummary} could be stored following an ARDM.
\par
The ARDM is an attempt to bring forward the problem of reproducibility and the lack of a single source of truth for analysis results. With it, we call for a paradigm shift where the target for the data analysis becomes the data model. Nonetheless, we understand the ARDM limitations and view it as one solution to a complex problem. We believe the best way to understand how the ARDM should evolve, or to shape it into a better solution, is to hear the opinions of the community. Hence, our underlying objective is to get the community's attention, discover similar initiatives, and converge on how to move forward in establishing analysis results as a data source to support future reusability and knowledge discovery.


\section*{Data availability}
The CDISC Pilot Project ADaM ADSL, ADTTE, and ADAE datasets were used to support the implementation of the analysis results data model. This data can be found at the PHUSE scripts repository \url{https://github.com/phuse-org/phuse-scripts/blob/master/data/adam/TDF_ADaM_v1.0.zip} and at the repository supporting this manuscript.


\section*{Code availability}
The implementation of the analysis results data model is available on Github (\url{https://github.com/joanacmbarros/analysis-results-data-model}). This repository exemplifies how to construct the data model and the respective schema, as well as show how to query the underlying database. Furthermore, we provide three output examples to visualize the results.


\bibliography{references}


\section*{Acknowledgements} 

We thank Carlotta Caroli, Nicholas Kelley, and Shahram Ebadollahi for their role in establishing and stewarding the AI4Life residency program. We also want to acknowledge Janice Branson for her valuable comments and support in this journey.


\section*{Author contributions statement}

All authors conceived and contributed to design the approach. M.B., L.A.W., and S.W. supervised the project. J.M.B. developed the data model and analysis standards. M.B. and L.A.W. reviewed the methodology. All authors read, edited, and approved the manuscript. 


\section*{Competing interests} 
The authors declare no competing interests  for this work.


\end{document}